%% file: main.tex
\documentclass[aps,prl,10pt,superscriptaddress,twocolumn]{revtex4-2}
\usepackage{blindtext}
\usepackage{amsmath,amsthm,amsfonts,amssymb,amscd}
\usepackage{verbatim}
\usepackage{mathrsfs}
\usepackage[svgnames,table,dvipsnames]{xcolor}
\usepackage{graphicx}
\usepackage{listings}
\usepackage{tikz}
\usetikzlibrary{decorations.text}
\usepackage{subfiles}
\usepackage{physics}
\usepackage{pgfplots}
\usetikzlibrary{arrows}
\usetikzlibrary{intersections, pgfplots.fillbetween}
\usepackage{epigraph} 
\usepackage{booktabs}
\usepackage[colorlinks=true,linkcolor=blue,allcolors=blue,citecolor=blue]{hyperref}
\usepackage{float}
\usepackage{mathtools}
\usepackage{caption}
\usepackage{subcaption}
\usepackage{bm}
\usepackage[capitalise]{cleveref}
\usepackage[export]{adjustbox}
\newcommand{\pd}{\partial}
\newcommand{\ep}{\epsilon}

\newcommand{\vel}{\bm{u}}
\newcommand{\velt}{u}
\newcommand{\rar}{\rightarrow}

\newcommand{\ci}{\mathrm{i}}
\newcommand{\taup}{{t'}}

\def\change#1{\textcolor{black}{#1}}
\definecolor{myblack}{RGB}{28,19,20}
\definecolor{myorange}{rgb}{1.0,0.55,0.0}
\definecolor{mydarkorange}{RGB}{152, 62, 25}
\definecolor{myred}{rgb}{0.55,0.0,0.0}

\makeatletter
\def\maketitle{
\@author@finish
\title@column\titleblock@produce
\suppressfloats[t]}
\makeatother

\begin{document}
\title{Breakup of an active chiral fluid}
\author{Luke Neville}
\email{lajn@mit.edu}
\affiliation{Department of Physics, Massachusetts Institute of Technology, Cambridge, Massachusetts 02139, USA}
\affiliation{School of Mathematics, Fry Building, University of Bristol, BS8 1UG, UK}
\affiliation{The Isaac Newton Institute for Mathematical Sciences, Cambridge CB3 0EH, UK}

\author{Jens Eggers}
\email{jens.eggers@bristol.ac.uk}
\affiliation{School of Mathematics, Fry Building, University of Bristol, BS8 1UG, UK}

\author{Tanniemola B. Liverpool}
\email{t.liverpool@bristol.ac.uk}
\affiliation{School of Mathematics, Fry Building, University of Bristol, BS8 1UG, UK}
\affiliation{The Isaac Newton Institute for Mathematical Sciences, Cambridge CB3 0EH, UK}

\begin{abstract}
The nonlinear breakup dynamics of a strip of active chiral fluid is considered, and it is shown that the strip thickness goes to zero as a power law in finite time. Applying slender body theory to the hydrodynamic equations of active chiral fluids, we predict the exponents analytically, and our predictions are shown to be in excellent agreement with numerical simulations. Qualitative agreement between experiment and simulation is also found.
\end{abstract}

\maketitle

\change{The decay of a fluid stream into drops is a highly nonlinear process, driven in passive systems by surface tension forces pulling in the stream from its surface. \cite{eggers1997nonlinear,eggers1993universal,eggersvillermaux,eggers2015singularities}. Little is known, however, when the fluid is active, or composed of individual agents that use energy to do work on their environment. These fluids, which show a variety of collective phenomena not seen in normal liquids \cite{ramaswamy2010mechanics,marchetti2013hydrodynamics}, differ from their passive counterpart by driving themselves unstable from the inside out \cite{marchetti2013hydrodynamics,zwicker2017growth,singh2019hydrodynamically,giomi2014spontaneous}. The effect of activity on interface dynamics has typically been studied in the linear or near-linear regimes. For example, small deviations from a flat interface \cite{adkins2022dynamics} or in active phase separation \cite{singh2019hydrodynamically,tjhung2018cluster} determining the long-time scaling and size-distribution of stable droplets. Here,  we look at the highly nonlinear dynamics near breakup of an unstable active film.}

% \change{The decay of a fluid stream into drops is a familiar process, and can be seen when water pours from a kitchen tap \cite{shi1994cascade,eggersvillermaux}, or when wax rises in a lava lamp. 

% % While the initial, linear, dynamics of both these processes look very similar, the nonlinear dynamics just before breakup look quite distinct \cite{eggers2015singularities}, with these differences arising from the different driving forces, like surface tension and inertia, that may dominate near breakup. Studying droplet formation thus lets us test the limit of different hydrodynamic theories by accessing their nonlinear regime.}

In particular, we examine a new type of fluid breakup where the instability is driven by the persistent spin of each constituent particle \cite{soni2019odd,liebchen2022chiral,han2021fluctuating,caporusso2024phase,furthauer2012activechiral,markovich2021odd}, which on hydrodynamic scales manifests itself as an anti-symmetric contribution to the stress tensor. Because the spin only affects the stress, these active chiral fluids behave identically to ordinary fluids in their bulk, with all effects of chirality propagating inwards from the boundaries \cite{ganeshan2017odd}. These effects were demonstrated in recent experiments \cite{soni2019odd} on a two-dimensional chiral fluid, made by sedimenting millions of particles onto a table and spinning them up with rotating magnetic fields \cite{massana2021arrested}. Taking thin strips of this chiral fluid, the authors of \cite{soni2019odd} observed counter-propagating flows along the strip boundaries, which eventually lead to breakup as the boundaries came into contact and sheared the fluid apart (see \cref{fig:experimental break and schematic}). This linear instability was then compared against the predictions from a hydrodynamic theory of chiral fluids, with excellent agreement found. The purpose of this letter is to go beyond the linear regime and study the fully nonlinear dynamics involved in breakup.

\begin{figure}[t]
\centering
\begin{subfigure}[t]{0.02\textwidth}
	\text{(a)}
\end{subfigure}\hfill
\begin{subfigure}[t]{0.45\textwidth}
  \centering
  \begin{adjustbox}{valign=t}
  \input figures/exp.tex
  \end{adjustbox}
\end{subfigure}\hfill
\begin{subfigure}[t]{0.02\textwidth}
	\text{(b)}
\end{subfigure}\hfill
\begin{subfigure}[t]{.45\textwidth}
  \centering
  \begin{adjustbox}{valign=t}
  \input figures/schematic_2.tex
\end{adjustbox}
\end{subfigure}
\caption{(a) Experimental evidence of two dimensional strips of chiral fluid breaking up asymmetrically, provided courtesy of the William Irvine group. The arrow marks the direction of time, with the strip going unstable before breaking up into drops. The whole process happens in a few seconds. (b) Sketch of a strip of chiral fluid undergoing an instability leading to breakup. The top and bottom surfaces are at $z=\pm h^{\pm}(x)$, the characteristic vertical scale is $h_0$, the horizontal scale is $L$, and their ratio is $\ep=h_0/L$. The dot-dashed orange line marks the centerline of the strip at $z=c(x)$, and the arrows inside the strip show the chiral shear flows}
\label{fig:experimental break and schematic}
\end{figure}

Our analysis uses asymptotics based upon slenderness to derive equations for the strip evolution. With these, the strip is shown to evolve self-similarly near breakup, with the minimum thickness decreasing to zero as a power law in finite time. The corresponding scaling exponent is anomalous \cite{barenblatt1996scaling,goldenfeld2018lectures}, and may not derived from dimensional considerations. Instead, it is found by solving a nonlinear eigenvalue problem that arises from a scaling analysis \cite{eggers2015singularities}, whose predictions perfectly match results from full PDE simulations.

\emph{Hydrodynamics.}--- Consider the strip of chiral fluid shown in \cref{fig:experimental break and schematic}. In the fluid bulk, inter-particle forces are balanced against the friction each particle feels from the the glass substrate. Using $\bm{\sigma}$ for the stress tensor and $\vel$ for the velocity, this gives
\begin{equation}
\div\bm\sigma-\Gamma\vel=\bm 0,
\label{eqn:stokes}
\end{equation} 
where $\Gamma$ is the friction coefficient of the substrate~\cite{soni2019odd,jia2022incompressible,ramaswamy1982linear}. \change{For chiral systems}, it is known \cite{soni2019odd,chaves2008spin,kirkinis2023activity}, that these hydrodynamic stresses can be described by a modified Newtonian stress tensor
\begin{equation}
\sigma_{ij}=\eta(\nabla_i \velt_j+\nabla_j \velt_i)-p\delta_{ij}+\eta_R\ep_{ij}(2\Omega-\omega),
\label{eqn:stress tensor}
\end{equation}
where $\eta$ is the dynamic viscosity, $p$ is the pressure \change{enforcing incompressibility}, and $\eta_R$  is the rotational viscosity \cite{furthauer2012activechiral}. The rotational viscosity term captures the friction felt whenever the spin rate of the particles, $\Omega$, is different to that of the surrounding water $\omega/2 = \ep_{ij}\nabla_i u_j/2$  \cite{chaves2008spin,kirkinis2023activity}.  In a passive system, conservation of angular momentum ensures the equality of the these spin rates and so the rotational viscosity term vanishes; \change{for the chiral systems considered here, $\Omega$ is constant and set by the external driving \cite{soni2019odd}}. Substituting the stress tensor (\ref{eqn:stress tensor}) into (\ref{eqn:stokes}) we find 
\begin{equation}
    (\eta+\eta_R)\nabla^2\vel - \bm{\nabla} p -\Gamma\vel=\bm{0},
\end{equation}
showing that chirality acts only to shift the viscosity in the fluid bulk. These chiral fluids therefore act the same as an ordinary fluids when surrounded by solid walls, as the no-slip boundary condition does not involve the particle spin, $\Omega$.

This spin, however, is picked up at free-surfaces, where the boundary condition, coming from a balance of stress and surface tension, is
\begin{equation}
\bm\sigma\cdot\bm{n}=-\gamma(\bm\nabla\cdot\bm{n})\bm{n}.
\label{eqn:dynamic}
\end{equation}
where $\bm{n}$ is the unit outward normal to the chiral fluid, $\gamma$ is the coefficient of surface tension \cite{landau1987fluid,zhang2020nanoscale}, and $\div\bm{n}$ is the mean curvature of the interface. The experimentally measured values of all parameters can be found in  \citet{soni2019odd}.

Using $(x,z)$ for coordinates in the down and cross strip direction, respectively, the top and bottom free surfaces have position $z=\pm h^{\pm}(x)$ (see \cref{fig:experimental break and schematic}(b)), and evolve in time according to the kinematic boundary condition
\begin{equation}
h^{\pm}_t+u h^{\pm}_x=\pm v\Big|_{z=\pm h^{\pm}},
\label{eqn:kinematic}
\end{equation}
where $(u,v)=\vel$, and subscripts denote derivatives. For a passive fluid we would expect the strip to evolve symmetrically with $h^+=h^-$, but chiral stresses break this symmetry. For later use we note that the height functions are related to the strip center-line and half-thickness by $c=(h^+-h^-)/2$ and $h=(h^+ + h^-)/2$, respectively. 

% \begin{table}
%    \centering
%    \begin{tabular}{*2c}
%        \toprule
%        Parameter & Experimental value\\
%        \midrule
%        Viscosity $\eta$ &  $4.9 \pm 0.2 \times 10^{-8}\ \mathrm{kg\ s^{-1}}$  \\
%        Rotational viscosity $\eta_R$ &  $9.1 \pm 0.1 \times 10^{-10}\ \mathrm{kg\ s^{-1}}$\\
%        Friction $\Gamma$ & $2.49 \pm 0.03 \times 10^{3}\ \mathrm{kg\ m ^{-2}\ s^{-1}}$ \\
%        Surface Tension $\gamma$ & $2.3 \pm 0.2 \times 10^{-13}\  \mathrm{kg\ m\ s^{-2}}$ \\
%        Rotation Rate $\Omega$ & $10-50\ \text{s}^{-1}$\\
%        \bottomrule
%    \end{tabular}
%    \caption{Experimentally measured values of the parameters appearing in the hydrodynamic equations. The rotation rate is varied in the experiments, and all values are taken from \citet{soni2019odd}. }
% \label{table: experimental values}
% %Odd Viscosity $\eta_0$ & $1.5 \pm 0.1 \times 10^{-8}\ \mathrm{kg\ s^{-1}}$ \\
% \end{table}

\begin{figure}[t]
\centering
\begin{subfigure}{0.03\textwidth}
	\text{(a)}
\end{subfigure}\hfill
\begin{subfigure}{0.435\textwidth}
  \centering
  \begin{adjustbox}{valign=t}
  \includegraphics[width=\textwidth]{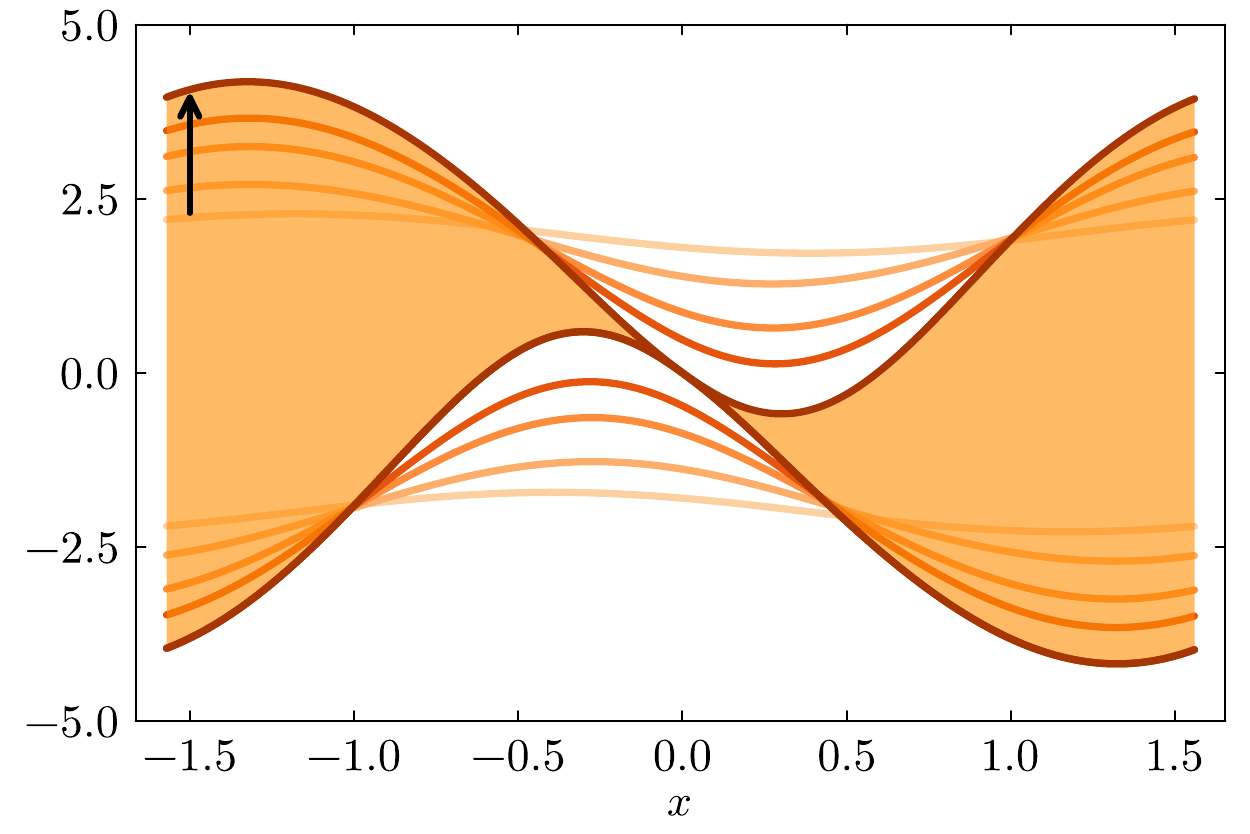}
  \end{adjustbox}
\end{subfigure}\hfill
\begin{subfigure}[t]{0.03\textwidth}
	\text{(b)}
\end{subfigure}\hfill
\begin{subfigure}{.45\textwidth}
  \centering
  \begin{adjustbox}{valign=t}
  \includegraphics[width=\textwidth]{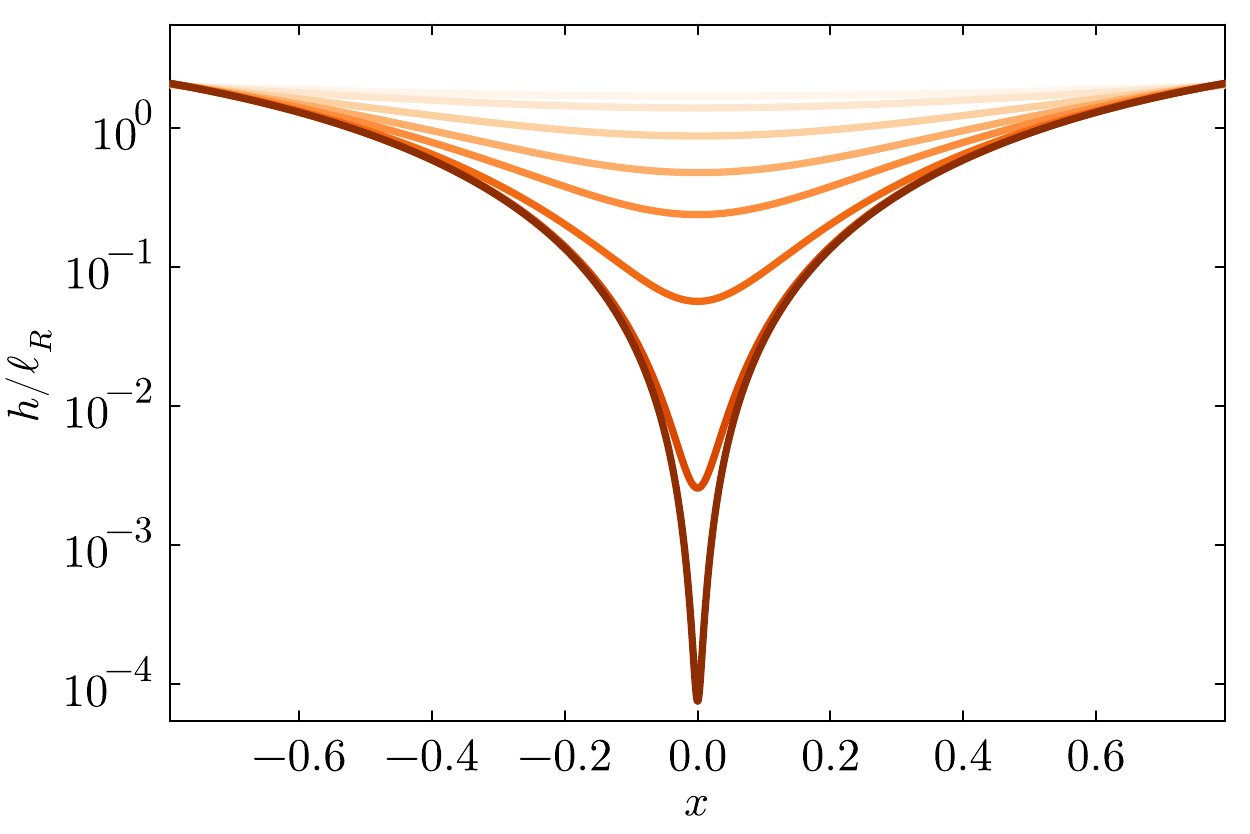}
\end{adjustbox}
\end{subfigure}
\caption{(a) Numerical solution of equations (\ref{eqn:dimfull 1d equations}) over a periodic domain. The light orange line shows the initial profile, and the sequence of darkening lines show the strip as time progresses; the black arrow marks the direction of time. (b) The thickness function $h$ is symmetric about $x=0$ as time progresses, and as breakup is approached the minimum strip thickness decreases to zero. In all plots lengths are measured in units of $\ell_R = \gamma(\eta+\eta_R)/(\eta\eta_R \Omega)$.}
\label{fig:pinch and logh}
\end{figure}

\emph{One-dimensional reduction.}--- To understand chiral breakup we must tackle the nonlinear dynamics of $h^{\pm}$ contained in equations (\ref{eqn:stokes}), (\ref{eqn:dynamic}), and (\ref{eqn:kinematic}), which can be difficult even with a sophisticated numerical method. We will take an alternative approach, exploiting the slenderness of the strip to project the dynamics down into one dimension \cite{eggers2015singularities,eggers1993universal,eggers1994drop,pdhowellsheet,papageorgiou1995breakup,erneux1993nonlinear}.

One-dimensional reductions have proven extremely successful in other problems \cite{eggers1994drop,zhang1999similarity,pdhowellsheet}, and follow the ideas of slender-body theory, expanding the dynamics in a small parameter given by the \change{ratio of the strips thickness to its horizontal length scale}. Leaving the derivation to the SI \cite{supplemental}, we find
\begin{equation}
\begin{gathered}
\Gamma c_t = -\frac{4\eta\eta_R \Omega}{\eta+\eta_R}\frac{h_x}{h} + \gamma \frac{c_{xx}}{h},\\
\Gamma h_t = -\frac{4\eta\eta_R \Omega}{\eta+\eta_R} (hc_x)_{xx} + \gamma(c_x c_{xx} -hh_{xxx})_x,
\end{gathered}
\label{eqn:dimfull 1d equations}
\end{equation}
which are two equations for the strip center-line and thickness. Keeping the parameters makes clear the origin of each term: those proportional to $\Omega$ come from the chiral stress, while others proportional to $\gamma$ come from surface tension. A linear stability analysis of Eqns. (\ref{eqn:dimfull 1d equations}) reproduces, at small initial thickness, the behavior of the full equations~\cite{soni2019odd}.
%To check our calculation we performed a stability analysis of a uniform strip, finding that Eqns. (\ref{eqn:dimfull 1d equations}) reproduced, for small wave-number, the stability analysis of the full equations (\ref{eqn:stokes}), (\ref{eqn:dynamic}), and (\ref{eqn:kinematic}) performed by \citet{soni2019odd}.

To go beyond the linear regime, we must integrate \eqref{eqn:dimfull 1d equations} numerically to understand nonlinearities. Our simulations were performed with periodic boundary conditions, using a well tested finite difference method~\cite{eggers2015singularities,eggers1994drop,bertozzi1994singularities}. 
Spatial derivatives were approximated to second order on a highly refined grid, that was adapted to scale with the pinch-region. Time integration was done using fully implicit step-halving method, ensuring both stability and second-order accuracy \cite{teukolsky1992numerical,supplemental}.

An example of a nonlinear evolution is shown in \cref{fig:pinch and logh}(a), where the initial shape is drawn in light orange, and a sequence of darkening lines shows the strip as time progresses. The final shape, which we shade in orange, pinches off at the origin. \change{Despite the profiles looking antisymmetric, the thickness function $h$ is symmetric} about the pinch-point, with all asymmetry in the strip coming from the totally anti-symmetric center-line. The global (anti-)symmetry of these shapes is a result of the initial conditions, but the symmetries were seen to hold true locally for all initial conditions we tried. The symmetry of $h$ is demonstrated nicely in \cref{fig:pinch and logh}(b), which shows a sequence of thickness profiles as the breakup time is approached.

\begin{figure}[t!]
\centering
  \includegraphics[width=0.48\textwidth]{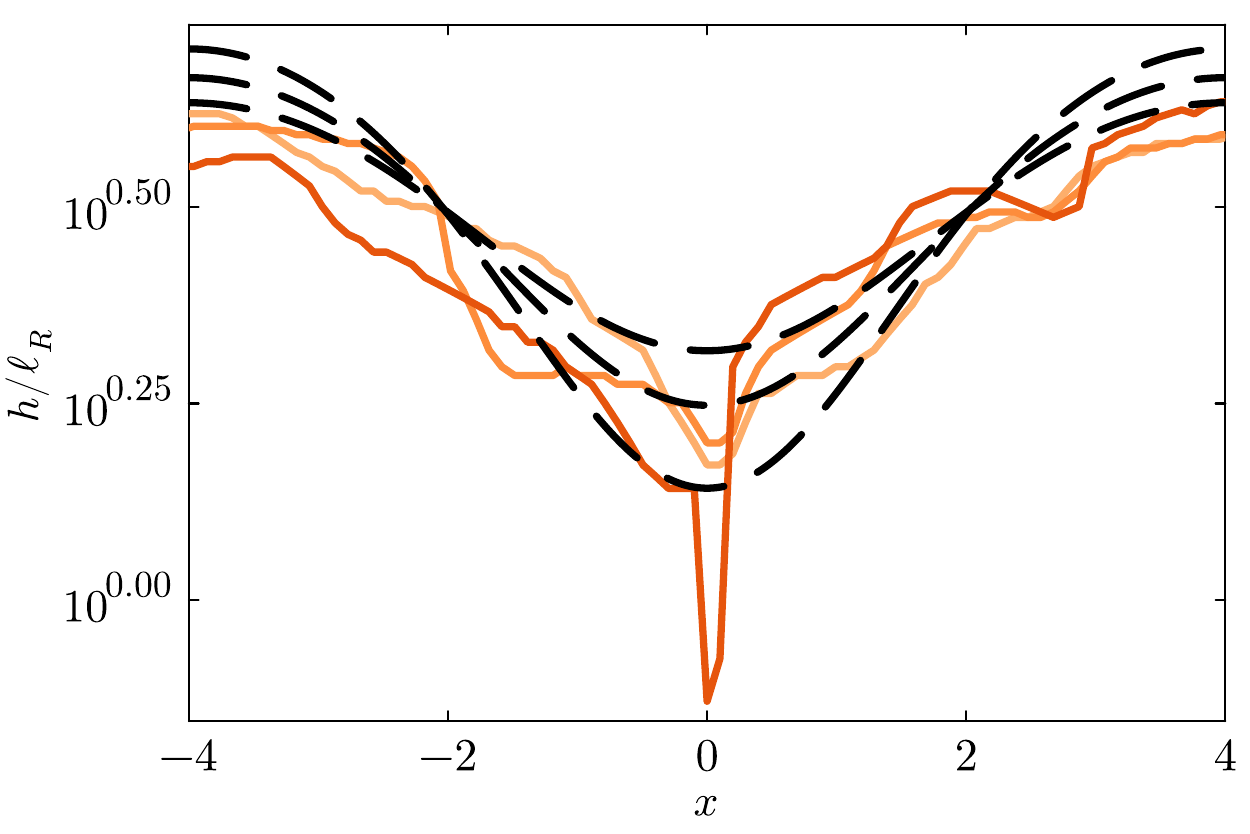}
\caption{A comparison between the experimental data from \cref{fig:experimental break and schematic}(a), shown in shades of orange, and numerical simulations, shown as dashed black lines. The simulations are initiated from an initial condition that smoothly approximated the experimental data, and each line in both sets are equally spaced in time. We see qualitative agreement between the experiment and simulation, with the timescale the only adjustable parameter.}
\label{fig:experiment compare}
\end{figure}

\change{To check our theory against experiment, we compare simulations run from the same, but smoothed, initial condition as the experiments shown in \cref{fig:experimental break and schematic}(a) with experiment. The results are shown in \cref{fig:experiment compare}, where the experiments, shown in shades of orange, match the simulations, shown as dashed black lines. Note that in comparing the results we have used the experimental values of the parameters taken from Soni et al. \cite{soni2019odd} to compute $\ell_R$, but adjusted the timescale. While there is not sufficient experimental data to follow the dynamics close to breakup, there is a clear transition between a linear (small deformation) and nonlinear regime.}

As the moment of breakup is approached, these nonlinearities become more important and control the dynamics. To understand them, we take a closer look at \cref{fig:pinch and logh}(b) to notice that the breakup is highly localized, with the thickness decreasing rapidly near the origin but changing slowly in the far-field. This suggests that a local analysis of the pinch region will be sufficient.

\emph{Scaling theory.}--- \change{Because the breakup is localized, its dynamics only depends on the length and time scales that are intrinsic to Eqns. (\ref{eqn:dimfull 1d equations})}, which are $\ell_R = \gamma(\eta+\eta_R)/(|\Omega|\eta\eta_R)$ and $t_R = \ell_R^3 \Gamma/\gamma$, respectively. Measuring dimensionless distances and times to the pinch-point at $(x_0,t_0)$ with $x'=(x-x_0)/\ell_R$ and $t'=(t_0-t)/t_R$, it is natural to assume that the system becomes self-similar as $t',x'\rar 0$ \cite{eggers2015singularities}, with
\begin{equation}
\begin{gathered}
h(x,t)=\ell_R \taup^{\alpha} f(x'/\taup^{\beta}),\\
c_x (x,t)=-S+\taup^{\alpha_2} g(x'/\taup^{\beta}).
\end{gathered}
\label{eqn:scaling form}
\end{equation}
The power law prefactors show that the strip thickness decreases to zero as ${t'}^{\alpha}$, with a universal shape given by the scaling function $f$. Following our simulation results we assume that the center-line slope tends to a constant value $-S$, with all spatial dependence sitting in a scaling function that rides on top \cite{eggers2000singularities}.

As we are interested in the final stages of breakup, we substitute (\ref{eqn:scaling form}) into (\ref{eqn:dimfull 1d equations}) and keep only the most dominant terms as  $t'\rar 0$.  A dominant balance argument \cite{bender2013advanced,eggers2015singularities} reveals these to be 
\begin{gather}
4f_\xi = g_\xi, \label{eqn:similarity equation g}\\
\frac{1+\alpha}{4}\xi f_{\xi}-\alpha f + (f f_{\xi\xi\xi})_\xi =0, \label{eqn:similarity equation f}
\end{gather}
where the similarity variable $\xi=x'/{t'}^{\beta}$, $\alpha_2=\alpha$, and $\beta=(1+\alpha)/4$. Any other choice of $\beta$ and $\alpha_2$ is inconsistent, \change{as other terms would become relevant} as $t'\rar 0$. The exponent $\alpha$ remains undetermined by dominant balance, \change{so we have self-similarity of the second kind} \cite{barenblatt1996scaling};  it is later found by solving the nonlinear eigenvalue problem (\ref{eqn:similarity equation f})  \cite{barenblatt1996scaling}.

The terms in (\ref{eqn:similarity equation g}) come from a balance of chirality and tension in the center-line equation; the time derivative term is negligible, and thus the center-line is quasi-static in the pinch region. The thickness equation, (\ref{eqn:similarity equation f}), comes from balancing the time derivative and the surface-tension terms in the thickness equation, and is locally equivalent to the Hele-Shaw equation when converted to similarity variables \cite{bertozzi1994singularities,eggers2008role,king2025touchdown}. However, in the Hele-Shaw problem, \change{a self-similar solution of the form (\ref{eqn:scaling form}) is never seen as the pinch region never becomes disconnected from the far-field \cite{dupont1993finite,almgren1996stable,bertozzi1994singularities}}. In contrast, here, the linear instability set in by the chiral stress is enough to kick-start breakup and we do see \change{ self-similarity} as in (\ref{eqn:scaling form}).

\begin{figure}[t]
\centering
\begin{subfigure}[t]{0.02\textwidth}
	\text{(a)}
\end{subfigure}\hfill
\begin{subfigure}[t]{0.46\textwidth}
  \centering
  \begin{adjustbox}{valign=t}
  \includegraphics[width=\textwidth]{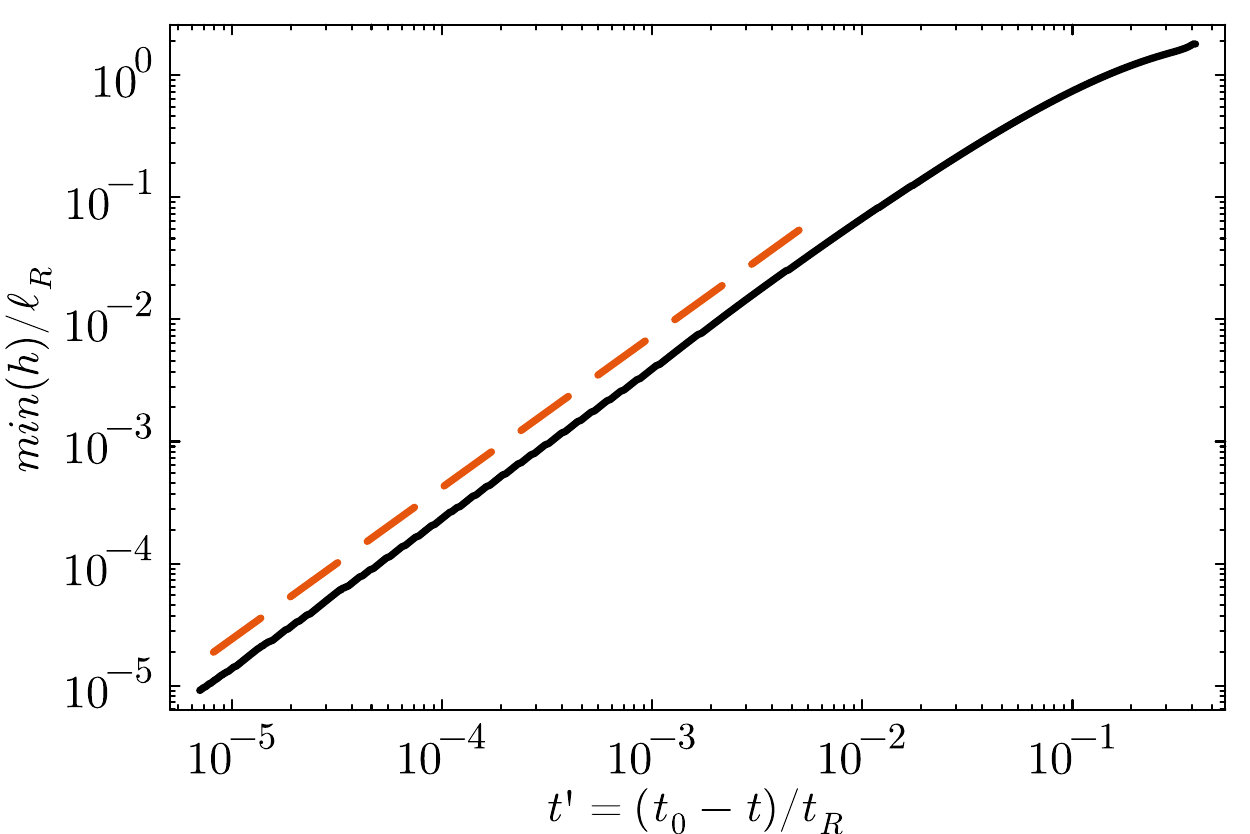}
  \end{adjustbox}
\end{subfigure}\hfill
\begin{subfigure}[t]{0.02\textwidth}
	\text{(b)}
\end{subfigure}\hfill
\begin{subfigure}[t]{.46\textwidth}
  \centering
  \begin{adjustbox}{valign=t}
  \includegraphics[width=\textwidth]{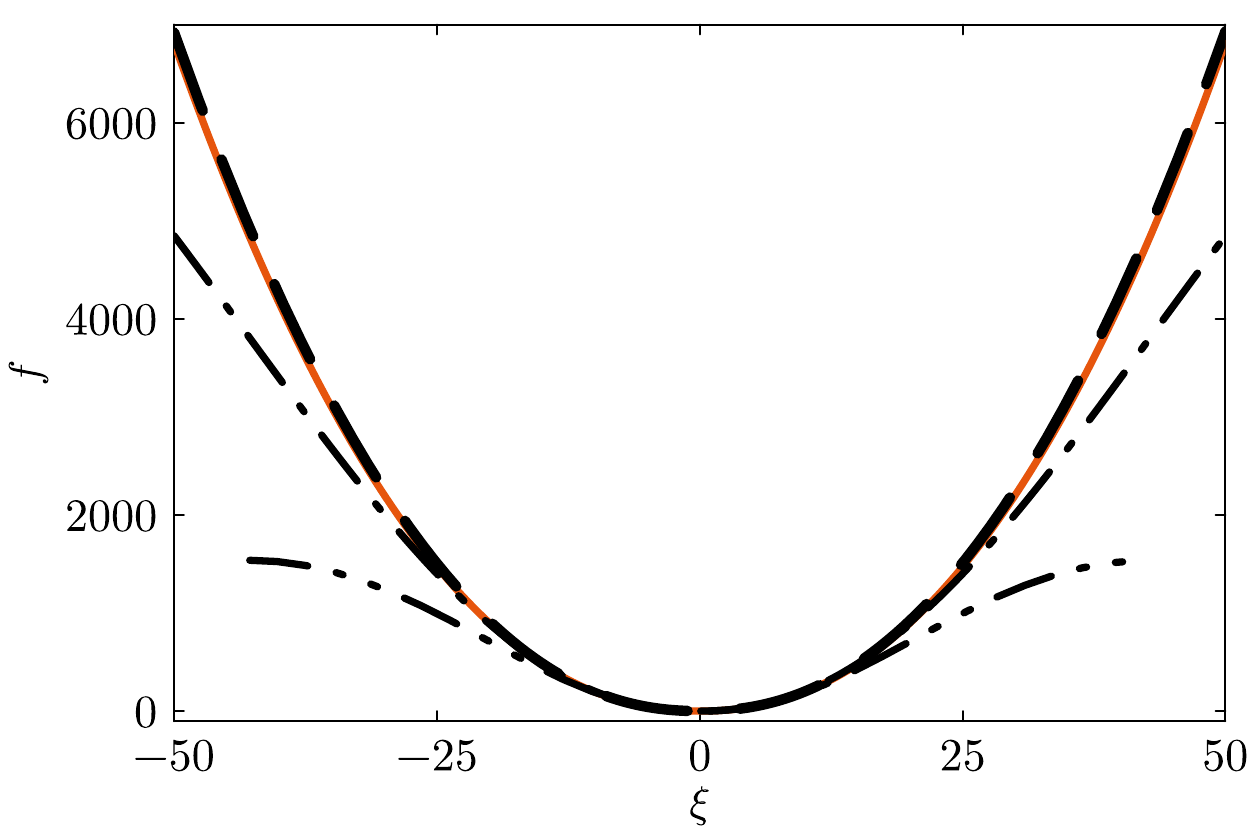}
\end{adjustbox}
\end{subfigure}
\caption{(a) Scaling of the minimum strip thickness as a function of the time to breakup, $t'$. The black line is from the PDE simulation, and the orange dashed line is the power law ${t'}^{1.24}$.  (b) The scaling function $f$. The solid orange line gives the prediction from our scaling theory; the dashed, dot-dashed, and chain-dashed lines come from the PDE simulation at $t'=7\times 10^{-6},1.8\times 10^{-4},7\times 10^{-4}$. The agreement between the PDE results and theory improves as breakup is approached.}
\label{fig:scaling and collapse}
\end{figure}

To find $g$ we integrate (\ref{eqn:similarity equation g}), giving
\begin{equation}
g = 4f+g_0,
\end{equation}
where $g_{0}$ is a constant, zero in all our simulations. The center-line and thickness are hence characterised by the same scaling function $f$, which solves (\ref{eqn:similarity equation f}). This equation is fourth order with one free parameter, meaning five boundary conditions are required for a unique solution. Assuming symmetry about the pinch at $\xi=0$ yields the first three
\begin{equation}
f(0)=1,\quad f_{\xi}(0)=0,\quad f_{\xi\xi\xi}(0)=0,
\label{eqn:symmetry conditions}
\end{equation}
where we have used a scale-invariance of (\ref{eqn:similarity equation f}) to set the value at the origin to unity. The remaining degrees of freedom are the exponent $\alpha$ and the second derivative at the origin $f_{\xi\xi}(0)$, which parametrise a two-dimensional space of solutions. A unique solution is picked out by enforcing the matching condition
\begin{equation}
f\propto \xi^{\alpha/\beta}\quad  \text{as}\quad |\xi| \rar\infty,
\label{eqn:matching condition}
\end{equation}
which ensures that the thickness becomes static far from the pinch region, with $t'$ dropping out of the scaling forms (\ref{eqn:scaling form}) as $\xi\rar\infty$ \cite{eggers2015singularities}. As shown in the SI \cite{supplemental}, matching corresponds to two conditions at infinity, and so (\ref{eqn:similarity equation f}) is now fully specified. The resulting problem is solved with the shooting method, yielding
\begin{equation} 
\alpha \approx 1.2392,\quad f_{\xi\xi}(0) \approx 2.7789,
\label{eqn:exponent values}
\end{equation}
as well as the solid red curve in \cref{fig:scaling and collapse}(b). Comparing this scaling with results from our PDE simulations in \cref{fig:scaling and collapse}(a) we see excellent agreement, with the strip thickness going to zero like $\taup^{1.24}$. The scaling function found from shooting also matches simulations, with the PDE results shown as dashed lines in \cref{fig:scaling and collapse}(b)  collapsing onto the predicted red curve as $t'\rar 0$. 

\change{The scaling exponent \eqref{eqn:exponent values} could not be extracted from the experimental data as there are not enough images during the breakup process \cite{soni2019odd}. To ease measurement of scaling, it would be beneficial to increase the values of $\ell_R$ and $t_R$, which would slow and enlarge the breakup region. This could be done by using particles with a stronger magnetic moment so that the surface tension was larger \cite{soni2019odd,chaves2008spin}.}

Our scaling analysis leads to a quantitative understanding of chiral breakup, which qualitatively can be understood by the following mechanism. The initial linear instability sets the system in motion and twists the center-line, resulting in a center-line-slope growing like $x^{k}$ for positive $k$, far away from the pinch. The balance of chiral and surface tension forces in (\ref{eqn:similarity equation g}) means that the thickness also grows, with $h\sim x^{k}/4$. Substituting this into (\ref{eqn:similarity equation f}) we see that the power law growth causes a surface-tension drive mass current that grows like $k(k-1)(k-2)x^{2k-3}$, pumping fluid out of the pinch provided that $k>2$. This agrees with the results from our scaling theory, as the matching condition (\ref{eqn:matching condition}) gives $k=\alpha/\beta\approx 2.2$. 

% \begin{figure}[t]
% \centering
% \includegraphics[width=0.48\textwidth]{figures/experiment_profiles_orange_log3.pdf}
% \caption{The thickness function extracted from the same system as \cref{fig:experimental break and schematic}(a). Each line is separated by 50 ms, and darkens in color as the breakup time is approached. As predicted from the theory, the profile is symmetric about the pinch (to within 10\% on average). }
% \label{fig:experiment shape}
% \end{figure}

To conclude, we have studied the asymmetric breakup of a strip of active chiral fluid using both asymptotics and numerics. Owing to novel features of the chiral dynamics, the strip exhibits a long anticipated similarity solution \cite{bertozzi1994singularities}, characterized by a set of universal scaling exponents, analogous to those found in critical phenomena \cite{goldenfeld2018lectures}. \change{ The asymptotic techniques used here could also be applied to other active matter problems where nonlinearities are relevant. Examples include the breakup or coalescence of active drops \cite{zwicker2017growth,singh2019hydrodynamically,soni2019odd}, and the spreading of active films \cite{zhao2024active}. }

\emph{Data Availability}.---The code used for numerical integration can be found at \cite{lukegithub} under ``chiral\underline{\hspace{.5em}}break".

\emph{Acknowledgements}.---TBL, LN would like to thank the Isaac Newton Institute for Mathematical Sciences, Cambridge, for support and hospitality during the programme {\em New statistical physics in living matter}, where part of this work was done. This work was supported by EPSRC grants EP/V520287/1, EP/R014604/1 and EP/T031077/1. We thank the group of William Irvine for supplying high resolution videos of the experiments. LN acknowledges the support of an EPSRC studentship and thanks Henry Andralojc for many interesting discussions, and  for help in understanding the symmetries of the problem.

\bibliography{chiral_biblio}

\clearpage

\subfile{chiral_supplementary.tex}

\end{document}

%% file: figures/exp.tex
 \begin{tikzpicture}[scale=1]
\node[] at (0,0)
    {\includegraphics[width=.7\textwidth]{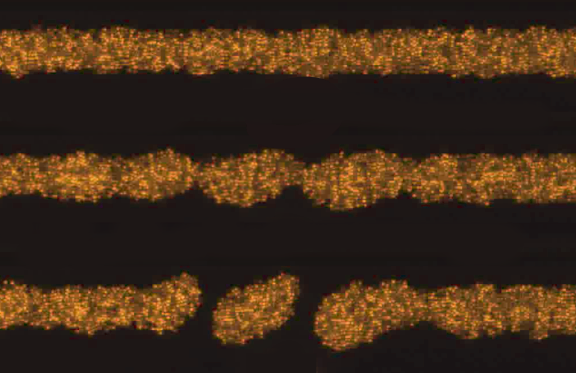}};
\draw[->,color=black,line width= .5mm] (-3.1,1.33) -- (-3.1,-1.2);
\node[rotate=90] at (-3.3,0) {time};
\node[text=white] at (2,.9) {$t=0s$};
\node[text=white] at (2,-.4) {$t=1.7s$}; %1.67 s%%
\draw [fill=myblack,myblack] (-2.815,-2) rectangle (2.817,-1.7); 
\node[text=white] at (2,-1.7) {$t=2.4s$}; % 2.35 s%
\end{tikzpicture}

%% file: figures/schematic_2.tex
\begin{tikzpicture}[scale=0.5,domain=-2:12]
\fill [white] (-3,3.2) rectangle (13,3.8);
\draw[->,color=black,line width= .5mm] (-3,0) -- (13,0);
\draw[->,color=black,line width= .5mm] (0,-2.5) -- (0,3);
\filldraw[color=black, fill=black,thick] (0,0) circle (.1);
\draw[name path= B, thin, mydarkorange,smooth]   plot (\x,{1.5+0.8*sin(16*3.14*\x)});
\draw[name path= A, thin, mydarkorange,smooth]   plot (\x,{-1.5-0.8*cos(16*3.14*\x)});
\tikzfillbetween[of=A and B]{myorange, opacity=0.6};
\draw[thick, mydarkorange,smooth, line width =.5mm]   plot (\x,{1.5+0.8*sin(16*3.14*\x)});
\draw[ thick, mydarkorange,smooth, line width =.5mm]   plot (\x,{-1.5-0.8*cos(16*3.14*\x)});

\draw[ thick, mydarkorange,smooth]   plot (\x,{-1.5-0.8*cos(16*3.14*\x)});

\draw[to-, line width=.3 mm, mydarkorange,smooth, domain = 10.5:12]   plot (\x,{1.5+0.8*sin(16*3.14*\x)});
\draw[to-, line width=.3 mm, mydarkorange,smooth, domain = 11:12]   plot (\x,{1.2+0.8*sin(16*3.14*\x)});
\draw[to-, line width=.3 mm, mydarkorange,smooth, domain = 11.5:12]   plot (\x,{.9 +0.8*sin(16*3.14*\x)});

\draw[-to, line width=.3 mm, mydarkorange,smooth, domain = 11.5:12]   plot (\x,{-.9-0.8*cos(16*3.14*\x)});
\draw[-to, line width=.3 mm, mydarkorange,smooth, domain = 10.8:12]   plot (\x,{-1.2-0.8*cos(16*3.14*\x)});
\draw[-to, line width=.3 mm, mydarkorange,smooth, domain = 10.5:12]   plot (\x,{-1.5-0.8*cos(16*3.14*\x)});

\draw[ thick, myred ,smooth,dash pattern={on 7pt off 2pt on 1pt off 3pt}]   plot (\x,{0.4*sin(16*3.14*\x)-0.4*cos(16*3.14*\x)});

\node[] at (0,3.2) {$z$};
\node[] at (12.9,0.5) {$x$};
\node[] at (10.8,2.6) {$z=h^+(x)$};
\node[] at (11,-1.7) {$z=-h^-(x)$};
\draw[<->,dashed,color=black,line width= .5mm] (3,1.8) -- (3,-0.7);
\node[] at (2.3,1.05) {$h_0$};
\draw[<->,dashed,color=black,line width= .5mm] (3.3,-2.7) -- (11,-2.7);
\node[] at (7.15,-3.1) {$L$};
\end{tikzpicture}

%% file: chiral_supplementary.tex
\title{Supplementary Material: Breakup of an active chiral fluid}

\maketitle

%%%%%%%%%% Merge with supplemental materials %%%%%%%%%%
%%%%%%%%%% Prefix a "S" to all equations, figures, tables and reset the counter %%%%%%%%%%
\setcounter{equation}{0}
\setcounter{figure}{0}
\setcounter{table}{0}
\setcounter{page}{1}
\makeatletter
\renewcommand{\theequation}{S\arabic{equation}}
\renewcommand{\thefigure}{S\arabic{figure}}

\onecolumngrid

\section{Deriving the equations}

To derive the one-dimensional equations from the full hydrodynamics we introduce characteristic horizontal and vertical scales $L$ and $h_0$, with ratio $\ep=h_0/L$. Assuming that this parameter is small, we non-dimensionalise all quantities according to 
\begin{equation}
\begin{gathered}
x\sim L,\quad z\sim \ep L,\quad h^{\pm}\sim \ep L,\\
u\sim u_0,\quad v\sim \ep u_0,\quad p\sim \frac{u_0(\eta+\eta_R)}{L},\\
\bm \sigma\sim \frac{\eta u_0}{L},\quad \gamma = \frac{\eta u_0}{\ep}\tilde{\gamma},\quad t\sim \frac{u_0}{L},
\end{gathered}
\end{equation}
where $u_0$ is a velocity scale to be set later. The scaling of surface tension is chosen to ensure it comes in at leading order \cite{pdhowellsheet}. Substituting these into the Stokes equation, we pick a vertical scale $h_0=\sqrt{(\eta+\eta_R)/\Gamma}$ to ensure that friction comes in to balance viscosity. The resulting equations are
\begin{subequations}
\begin{gather}
(\ep^2 \pd_x^2 +\pd_z^2)u -\ep^2 \pd_x p - u=0,\\
(\ep^2\pd_x^2 +\pd_z^2)v -\pd_z p - v=0,\\
\pd_x u +\pd_z v=0,
\end{gather}
\label{eqn:scaled stokes}
\end{subequations}
where all variables and coordinates are now dimensionless.

Assuming that the velocity scale is set by the rotation rate of the cubes we choose $u_0=\Omega h_0$, yielding non-dimensionalised stress components
\begin{equation}
\begin{gathered}
\sigma_{xx}=2\pd_x u -p(1+R),\\
\sigma_{zz}=2\pd_z v -p(1+R),\\
\sigma_{xz}=\ep^{-1}(1+R)\pd_z u +\ep(1-R)\pd_x v-2R\ep^{-1},\\
\sigma_{zx}=\ep^{-1}(1-R)\pd_z u +\ep(1+R)\pd_x v-2R\ep^{-1}.
\end{gathered}
\label{eqn:scaled stress}
\end{equation}
where $R=\eta_R/\eta$ is the ratio of the rotational and dynamic viscosities. The dimensionless dynamic boundary condition is given by
\begin{equation}
\ep \bm\sigma\cdot\textbf{n}^{\pm}=-\tilde{\gamma}(\nabla\cdot\textbf{n}^{\pm})\textbf{n}^{\pm},
\label{eqn: scaled free surface}
\end{equation}
while the kinematic condition remains the same.

We solve equations (\ref{eqn:scaled stokes}) perturbatively in the small parameter $\ep$, expanding the fields as a series in $\ep^2$,
\begin{subequations}
\begin{gather}
\textbf{u}=\vel_0+\ep^2 \vel_1+...,\\
p=p_0+\ep^2 p_2+...,
\end{gather}
\end{subequations}
while assuming $h^{\pm}$ are $O(1)$. At leading order for $u_0$ we have 
\begin{subequations}
\begin{gather}
\pd_z^2 u_0 -u_0,\\
\pd_z u_0 \Big|_{z=\pm h^\pm}=\frac{-2R}{1+R},
\end{gather}
\end{subequations}
where the boundary condition comes from the balance of tangential stresses along the free surface. Solving these gives the zeroth order velocity as
\begin{equation}
u_0=-\frac{2R}{1+R}\sech\left(h\right)\sinh(z-c).
\end{equation}
Combing this with the incompressibility condition, the other component of the Stokes equation and the normal stress boundary conditions, we can compute $v_0$ and $p_0$. The results are quite lengthy, but simplify tremendously when substituted into the kinematic boundary conditions (\ref{eqn:kinematic}). The resulting equations of motion for the center-line and thickness are
\begin{equation}
\begin{gathered}
h_t=0,\\
c_t = -\frac{4R \sech(h)^2 h_x}{(1+R)^2 h}+\frac{\tilde{\gamma} c_{xx}}{(1+R)h},
\end{gathered}
\label{eqn:leading order thickness}
\end{equation}
where subscripts denote derivatives. The thickness remains quasi-stationary at this order, which we soon see is because its equation is higher order in a derivative expansion. Pushing the calculation to next order in $\ep$ is conceptually easy, but technically difficult due to the complicated algebra. The next-order velocities and pressures were computed with mathematica and take many pages. Thankfully they simplify when plugged into the kinematic condition (\ref{eqn:kinematic}) to give
\begin{equation}
h_t = \ep^2 \Bigg[-\frac{4R c_x\tanh(h)}{(1+R)^2}+\frac{ \tilde{\gamma}\left(c_x^2+h_x^2-2 h h_{xx}\right)}{2(1+R)}\Bigg]_{xx},
\label{eqn: higher order thickness}
\end{equation}
with the dynamics of the center-line given by (\ref{eqn:leading order thickness}). The two equations (\ref{eqn:leading order thickness}) and (\ref{eqn: higher order thickness}) constitute a one-dimensional reduction of the full equations (\ref{eqn:stokes}), (\ref{eqn:dynamic}), and (\ref{eqn:kinematic}), valid when the strip is slender. The equations of the main text are found by expanding the hyperbolic functions for small $h$, which is valid when the strip is thin, and then returning back to dimension-full coordinates.

We have checked that simulations of the full equations, including the hyperbolic functions obey the same scaling relations as the simplified equations from the main text.

\section{Numerical integration}

For our numerical work we use a fully implicit finite difference method. The simulation domain was chosen to be periodic and of length $L$, with grid points at positions $x_i$ and grid spacings $\Delta x_i = x_{i+1}-x_i$. Spatial derivatives were approximated to second order accuracy using the well-known Fornberg algorithm \cite{fornberg1988generation}, transforming, the partial differential equations into a set of difference differential equations.

Because of the separation in timescale between the slow dynamics far from breakup, and the fast ones near it, the equations are stiff. As such, it was beneficial to use an implicit time integrator to ensure stability without the need of a very small time step. The equations were written implicitly as 
\begin{equation}
    \frac{\bm{F}_{n+1} - \bm{F}_{n}}{\Delta t} = \mathcal{N}[\bm{F}_{n+1}],
\end{equation}
where $\bm{F}$ is the vector of $h,c$ values at each grid point, $\Delta t$ is the time step, $n$ is the current time, and $\mathcal{N}$ is a nonlinear function containing all the right-hand terms. To solve these equations we used Newton's method, taking advantage of the sparsity of the Jacobian to speed up computations, and using a forward Euler step as an initial guess. 

To improve the accuracy of this method we used a step halving method, combining a solution from one step of size $\Delta t$ and two steps of size $\Delta t/2$ to arrive at a solution that is accurate to second order in $\Delta t$ \cite{eggers2015singularities}. The time step was adapted to ensure that the numerical error remained low, and that $\text{min}(h)$ decreased by less than $5\%$ in each step. 

The spatial grid was also adapted to match the scale of the solution, with the grid being fine and constant near the pinch point, and increasing to a larger value away from it. Every time the minimum thickness decreased by $10\%$ we interpolated the solution onto a new grid with a finer resolution and kept integrating. This adaptability allowed us to follow the dynamics to very small scales without the computational cost that would come from a very fine uniform grid.

\section{Stability and selection of the similarity equations}

In the main body of the text we stated that the solution depended on two parameters $\alpha$ and $f_{\xi\xi}(0)$, and that they had to be tuned to satisfy the matching condition $f\propto \xi^{\alpha/\beta}$ at infinity. The dependence on two parameters follows from calculating the power series expansion of $f$ near $\xi=0$. I.e. we substitute the expansion
\begin{equation}
f=1+\sum_{n=1}^{\infty} a_n \xi ^{2n},
\end{equation}
into the equation and solve for all the coefficients $a_n$. This can be done to arbitrarily high order using, say, mathematica, and it shows that all the coefficients depend on $\alpha$ and $a_1=f_{\xi\xi}(0)/2$. For example, the first two coefficients are
\begin{equation}
a_2 = \alpha/24,\quad a_3 = -a_1(1+5\alpha)/720.
\end{equation}
To demonstrate that a tuning of the two parameters amounts to a selection of the right solution we now perform a stability analysis, perturbing around the far-field behaviour by setting
\begin{equation}
f=\xi^{\alpha/\beta}+\delta P(\xi),
\end{equation} 
and keeping terms linear in $\delta$. The perturbations $P$ will fall  into two classes: unstable perturbations that grow faster than $\xi^{\alpha/\beta}$ at infinity, and stable perturbations that are constant or decay \cite{eggers2015singularities}. Although the perturbations $P$ cannot be found exactly, a WKB analysis is sufficient to understand the stability at infinity. Upon substitution of $P=e^{\chi}$ into the equation, we keep the most dominant terms as $\xi\rar\infty$ to get
\begin{equation}
\frac{1+\alpha}{4}\xi \chi' = -\xi^{\alpha/\beta}(\chi')^4,
\end{equation}
where $'=\pd_{\xi}$. Solving this equation we find four modes $P_i =e^{\chi_i}$, where
\begin{gather}
\chi_1\sim\text{const.},\quad \chi_2 \sim -\frac{3}{4}\left(\frac{1+\alpha}{4}\right)^{4/3}\xi^{4/(3+3\alpha)},\\
\chi_3 \sim \frac{3}{8}\left(\frac{1+\alpha}{4}\right)^{4/3}\xi^{4/(3+3\alpha)}(1+\sqrt{3}\ci),\quad \chi_4 \sim \frac{3}{8}\left(\frac{1+\alpha}{4}\right)^{4/3}\xi^{4/(3+3\alpha)}(1-\sqrt{3}\ci).
\end{gather}
The first mode $P_1$ corresponds to a shift in the parameters of the solution. The second mode $P_2$ is exponentially decaying at infinity and is stable. The remaining modes $P_3$ and $P_4$ are oscillatory, exponentially growing modes whose prefactors must be tuned to zero to hit the matching condition. Because the whole solution depends only on $\alpha$ and $a_1$, this tuning is enough to select the solution in the main text. 

Unfortunately the exponential growth is extremely weak, being approximately $e^{\xi^{0.6}}$. Numerically, this means we have to shoot to quite large values of $\xi$ to attain convergence in $\alpha$ and $a_1$.